\documentclass{article}

\usepackage[english]{babel}

\usepackage[letterpaper,top=2cm,bottom=2cm,left=3cm,right=3cm,marginparwidth=1.75cm]{geometry}

\usepackage{amsmath}
\usepackage{amssymb}
\usepackage{graphicx}
\usepackage[colorlinks=true, allcolors=blue]{hyperref}
\usepackage{color,soul}
\usepackage{comment}

\title{

Detection of spiking motifs of arbitrary length in neural activity using bounded synaptic delays
}
\author{T. Kronland-Martinet \and S. Viollet \and L. Perrinet}

\begin{document}
\maketitle

\begin{abstract}
In the context of spiking neural networks, the temporal coding hypothesis is increasingly preferred over the rate coding hypothesis due to its advantages in processing speed and energy efficiency. In temporal coding, synaptic delays are crucial for processing signals with precise spike timings, known as spiking motifs. Synaptic delays are however bounded in the brain and can thus be shorter than the duration of a motif. This prevents the use of motif recognition methods that consist of setting heterogeneous delays to synchronize the input spikes on a single output neuron acting as a coincidence detector. 
To address this issue, we developed a method to detect motifs of arbitrary length using a sequence of output neurons connected to input neurons by bounded synaptic delays. Each output neuron is associated with a sub-motif of bounded duration. A motif is recognized if all sub-motifs are sequentially detected by the output neurons. We simulated this network using leaky integrate-and-fire neurons and tested it on the Spiking Heidelberg Digits (SHD) database, that is, on audio data converted to spikes via a cochlear model, as well as on random simultaneous motifs.
The results demonstrate that the network can effectively recognize motifs of arbitrary length extracted from the SHD database. Our method features a correct detection rate of about 60\% in presence of ten simultaneous motifs from the SHD dataset and up to 80\% for five motifs, showing the robustness of the network to noise. Results on random overlapping patterns show that the recognition of a single motif overlapping with other motifs is most effective for a large number of input neurons and sparser motifs. Our method provides a foundation for more general models for the storage and retrieval of neural information of arbitrary temporal lengths.
\end{abstract}

\section{Introduction}

The pioneering work of Mainen and Sejnowski~\cite{mainen1995reliability}, in 1995, demonstrated that neurons can generate reproducible spikes with sub-millisecond precision.
Since then, increasing evidence has shown that such precise spike timings play a crucial role in neuronal communication within the brain~\cite{grimaldi2022precise}.
The temporal coding hypothesis has gained significant attention due to its efficiency in processing speed compared to rate coding, which typically requires longer processing times~\cite{van2001rate, brette2015philosophy}. 
A crucial model was found in spiking neural networks where neurons are randomly connected with random frozen synaptic delays~\cite{izhikevich2006polychronization}. Groups known as polychronous groups defined by precise millisecond-scale spiking motifs have been observed to naturally emerge as a consequence of Spike-Timing Dependent Plasticity (STDP) learning~\cite{izhikevich2004spike}. These polychronous groups were reoccurring multiple times during simulations, even when the network inputs were random, suggesting that learning motifs with precise spike timings could be highly relevant in the brain~\cite{villette_internally_2015}.

Moreover, reducing time precision leads to a degradation of results such that the precise timing of spikes is crucial for motif recognition tasks~\cite{akolkar2015can}. Previous studies on motif recognition have primarily focused on synaptic weights, as seen in Hopfield networks~\cite{hopfield1982neural}, often disregarding the temporal precision of memorized motifs. Some Hebbian learning rules, such as STDP, consider the precise timing of spikes for learning but most often update only the synaptic weights~\cite{Perrinet2002b}.
Other studies have suggested strategies for recognizing temporal motifs. For instance, networks using bistable neurons and global inhibition can recognize sequences of spikes of arbitrary length~\cite{yu2015spiking, jin2004spiking, jin2008decoding}. However, these networks do not account for precise spike timings. In contrast, networks with heterogeneous delays have been used to recognize motifs with precise spike timing~\cite{unnikrishnan1991connected, tank1987neural}. However, although delays can be set with arbitrary lengths on a computer, they are constrained in the brain, with synaptic delays generally considered to be shorter than 30~ms~\cite{Swadlow1985,lemarechal2022brain}. Therefore, as defined in~\cite{unnikrishnan1991connected, tank1987neural}, synaptic delays highly constrain the length of stored motifs. Learning rules, such as the DELTRON~\cite{hussain2012deltron}, have been developed to learn delays but do not consider the limited range of delays imposed by biological neurons. 

To overcome these constraints, we developed a network architecture capable of detecting motifs of arbitrary length using bounded delays. This network operates by dividing a spiking motif of arbitrary length into sub-motifs of bounded duration. For each sub-motif, an output neuron is connected to the input neurons to retain the sub-motif. Output neurons are then connected together sequentially, and a motif is recognized if the final output neuron fires following the sequential activation of all sub-motifs. We tested this network in simulations using audio data converted to spikes via a model of the cochlea. The network successfully recognized motifs even when they were presented simultaneously. Additionally, we evaluated the network's capacity to recognize a motif in the presence of simultaneous motifs using random synthetic motifs. This test, related to the cocktail party problem~\cite{mcdermott2009cocktail}, highlighted the network's ability to recognize motifs even with high levels of background noise.
In the following sections, we first describe the network in Section~\ref{section:motif_recognition_network}, then detail the simulation process in Section~\ref{section:Simulation of the network}, and analyze the results in Section~\ref{section:results}. Finally, we discuss the results in Section~\ref{section:discussion}.

\section{Architecture of a recognition network for arbitrary-length motifs}
\label{section:motif_recognition_network}
\subsection{Network architecture}
\label{section:network_architecture}

In this section, we describe the architecture of the network designed to recognize a single spiking motif generated by a group of input neurons. Recognition is achieved through the sequential activation of output neurons, each corresponding to a sub-motif, that is, a segment of the motif. Figure~\ref{fig:neural_architecture} provides a visual overview of this architecture.

\subsubsection{Definition of spiking motifs}

\begin{figure}[htbp]
    \centering
    \includegraphics[width=0.6\textwidth]{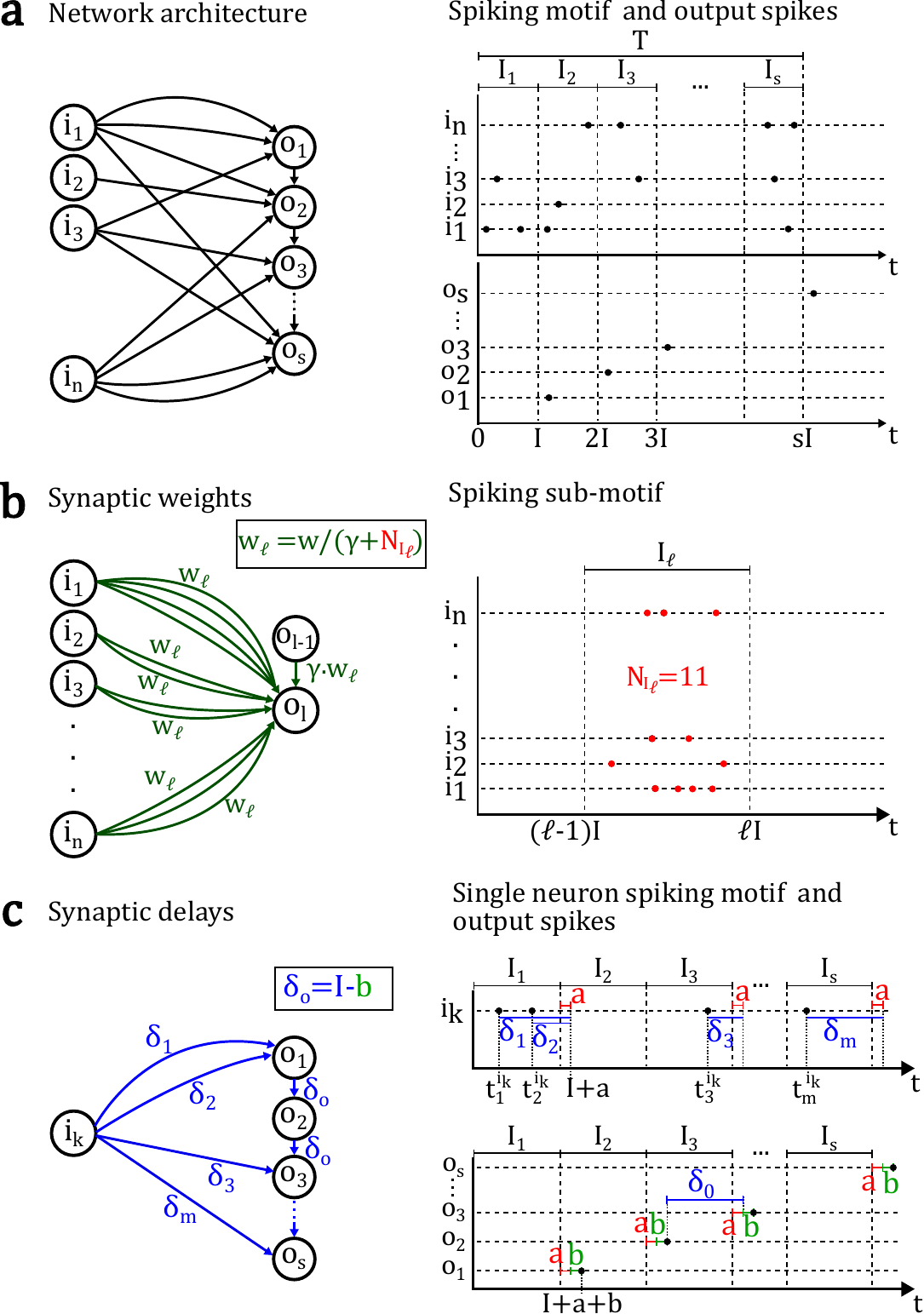}
    \caption{Neural network architecture for the recognition of arbitrary-length motifs (each black or red dot on right-side figures correspond to a spike). a) Network achitecture : the connectivity between input and output neurons follows the distribution of input spikes within each sub-motif $I$. b) Distribution of the synaptic weights according to equation \ref{eq:weight_input_output_factor}. c) Calculus of the synaptic delay according to equations~\ref{eq:delay_input_output} and \ref{eq:delay} and their representation for a given motif.}
    
    \label{fig:neural_architecture}
\end{figure}

Let us consider $n\in\mathbb{N}^*$ input neurons that fire a motif of duration $T\in\mathbb{R}^{+*}$ \footnote{We define $\mathbb{N}^*$ as the set of non zero integers, while $\mathbb{R}^{+*}$ is the set of strictly positive real numbers.}.
The input neurons are labelled $i_1, i_2,..., i_n$ and a neuron $i_j$, with $j \in \{1,...,n \}$, emits  $m \in \mathbb{N}$ spikes with timings $\{ t^{i_j}_{1}, t^{i_j}_{2}, ...,  t^{i_j}_{m} \} \in \mathbb{R}^m$. These timings are ordered, that is, $0 \leq t^{i_j}_{1} < t^{i_j}_{2} < ... < t^{i_j}_{m} < T$ (cf. figure~\ref{fig:neural_architecture}c).
The motif is defined as a sequence of spikes from the input neurons, where the timing of each spike is precise. The goal of the network is to recognize this motif by detecting the precise timing of spikes in the input neurons.

To construct the network, the motif duration $T$ is divided in $s\in\mathbb{N}^*$ intervals of length $I= \frac T s$ (cf. figure~\ref{fig:neural_architecture}a). The parameter $s$ should be chosen such that the length of the intervals is shorter 
than the maximum possible synaptic delay. We can compute the number of all input neurons' spikes in the motif that occurs in each interval $I_\ell = \left[ \left(\ell-1 \right)I, \ell I \right[$, where $\ell \in \{1,...,s\}$, as follows:
\begin{equation}
    N_{I_\ell} = \sum_{j =1}^{n} \sum_{k =1}^{m} \mathbf{1}_{\left(\ell-1\right)I\leq t^{i_j}_k<\ell I}
\end{equation}
Where $\mathbf{1}_{C}$ is the Kronecker delta function, which is equal to 1 if the condition $C$ is true and 0 otherwise (cf. figure~\ref{fig:neural_architecture}b). 

\subsubsection{Sub-motif detection}
\label{section:Sub-motif detection}
The input neurons are connected to a group of $s$ output neurons labelled $o_1, o_2,..., o_s$ each associated to a time interval. Each output neuron is such that $o_\ell$ has $N_{I_\ell}$ synapses from input neurons. For each spike timing $t^{i_j}_k$ such that $\left(\ell -1\right)I\leq t^{i_j}_k<\ell I$, a synapse connects input neuron $i_j$ to output neuron $o_\ell$ with a delay tuned such that the spikes converging on the soma of output neuron $o_\ell$ synchronize to the end of the interval $I_\ell$. In other words, we define the synaptic delay $\delta_{i_j,o_\ell,k}$ from neuron $i_j$ to output neuron $o_\ell$ such that if neuron $i_j$ spikes at time $t^{i_j}_k$, with respect to the beginning of the motif, this spike reaches neuron $o_\ell$ at time $t_{i\ell} = \ell I+a$. Here, $a > 0$ is a constant added to ensure that every synaptic delay is strictly positive, i.e., $\delta_{i_j,o_\ell,k} \geq a$. This prevents any synapse from having zero delay, which is biologically implausible and problematic in simulations. As a consequence, if delays are attuned regarding the following equation : 
\begin{equation}
    t_{i\ell} = \ell I + a = \delta_{i_j,o_\ell,k} + t^{i_j}_k
\end{equation}
then all spikes from the sub-motif reach the output neuron at the same precise time. 
Finally, the synaptic delay is therefore defined as (cf. figure~\ref{fig:neural_architecture}c) :
\begin{equation}
    \delta_{i_j,o_\ell,k} = \ell I + a - t^{i_j}_k
    \label{eq:delay_input_output}
\end{equation}

The synaptic weights are defined such that neuron $o_\ell$ fires only if all spikes of the motif occurring in the interval $I_\ell$ reach simultaneously the neuron $o_\ell$. Moreover, neurons operate in the coincidence detector regime~\cite{konig1996integrator}, so that only recent input spikes contribute significantly to spike generation. Details are given in the following section. 

\subsubsection{From local to global detection}

An input spiking motif may be trigered after a silent period, such that we define the first output neuron $o_f$ having connections with input neurons by its index $f$:   
\begin{equation}
    f = \min_\ell \left( N_{I_\ell} >0 \right)
\end{equation}
For output neuron $o_f$, the synaptic weights are defined such that $o_f$ fires only if it receives $N_{I_f}$ synchronous spikes. Let us consider $w$ the minimum weight such that a Leaky Integrate-and-Fire (LIF) neuron fires from a single input  (see next section for more details). The synaptic weight associated with each input synapse to output neuron $o_f$ is given by 
\begin{equation}
    w_f = \frac{w}{N_{I_f}}
\end{equation}
To allow for noisy inputs, a threshold can be defined such that $o_f$ fires if input spikes are missing or if the input spikes are only approximately synchronous. Equivalently, we can define a new weight: 
\begin{equation}
    w_{f,noisy} = w_f + \epsilon, \epsilon \in \mathbb{R}^+
    \label{eq:noisy_input_weight}
\end{equation}

To ensure a chain of detection of the global motif by the convergence of the spiking motifs within each interval $I_\ell$, each output neuron $o_\ell$ with $f\leq \ell<s$ is connected to the next output neuron $o_{\ell+1}$ with a synaptic delay $\delta_{o_\ell,o_{\ell+1}}$. As specified in section \ref{section:Sub-motif detection}, if the memorized motif occurs, the output neuron $o_\ell$ should receive all simultaneous input spikes at time $t_{i\ell} = \ell I + a$, with respect to the beginning of the motif. Therefore, by considering its dynamic, neuron $o_\ell$ would fire at time $t_{o\ell} = \ell I + a + b$, where $b>0$ is the the delay neuron $o_\ell$ needs to generate a spike when receiving the stored input signal. Similarly, output neuron $o_{\ell+1}$ should receive all input spikes at time $t_{i(\ell+1)} = (\ell+1) I + a$, including the spike from output neuron $o_\ell$. Therefore, the delay $\delta_{o_\ell,o_{\ell+1}}$ is defined such that (cf. figure~\ref{fig:neural_architecture}c):
\begin{equation}
\begin{aligned}
    &t_{i(\ell+1)} = t_{o\ell} + \delta_{o_\ell,o_{\ell+1}} 
    \\ \iff &(\ell+1) I + a = \ell I + a + b + \delta_{o_\ell,o_{\ell+1}} 
    \\ \iff &\delta_{o_\ell,o_{\ell+1}} = I - b
\end{aligned}
\label{eq:delay}
\end{equation}

Except for output neuron $o_f$, the synaptic weight of input synapses to output neuron $o_\ell$ are given by:
\begin{equation}
    w_\ell = \frac{w}{N_{I_\ell}+1}
\end{equation}
Such that neuron $o_\ell$ spikes only if it receives $N_{I_\ell}$ synchronous spike from input neurons and one more synchronous spike from the previous output neuron.

If we want to ensure that the output neurons spike only if the previous output neuron spiked, we can define the synaptic weight of input synapses to output neuron $o_\ell$ by (cf. figure~\ref{fig:neural_architecture}b):
\begin{equation}
    w_\ell = \frac{w}{N_{I_\ell}+\gamma}
    \label{eq:weight_input_output_factor}
\end{equation}
In that equation, $\gamma>1$ is a parameter giving the relative strength of output to output synapses with respect to input to output synapses, such that: 
\begin{equation}
    w_{o_{\ell-1}o_{\ell}} = \frac{\gamma w}{N_{I_\ell}+\gamma} = \gamma w_\ell
    \label{eq:weight_output_output_factor}
\end{equation}
is the weight between two successive output neurons $o_{\ell-1}$ and $o_{\ell}$. By increasing the value of $\gamma$, we reduce the probability for neurons to spike independently from the memorized sequence.

All weights can be adjusted to process noisy motifs, as shown in eq.~\ref{eq:noisy_input_weight}. Finally, the motif is recognized if and only if neuron $o_s$ fires.

\subsection{Network implementation using LIF neurons}

\subsubsection{Definition of $w$ and $b$ from the Leaky Integrate-and-Fire (LIF) neuron equation}

To explain the choice of parameters $w$ and $b$ defined in the previous section, we consider the LIF model of a spiking neuron for its simplicity and because it was sufficient for the efficient simulation of the network.
A LIF neuron's membrane potential is defined by~\cite{gerstner2014neuronal}:
\begin{equation}
    \tau_m\frac{du}{dt} = -\left[u(t)-u_\text{rest}\right]+RI(t)\label{eq:LIF_membrane_potential}
\end{equation}
Where $u$ is the membrane potential of the neuron, $\tau_m$ its membrane time constant, $u_\text{rest}$ is the neuron's resting membrane potential, $R$ the membrane resistance and $I$ the input signal's intensity.
\\
The solution of the LIF membrane potential equation for a neuron receiving an input spike of intensity $I(t)=w I_0$ for $0<t<\Delta$ is: 
\begin{equation}
    \begin{cases}
        u(t) = u_\text{rest} &\quad \text{if } t\leq 0 \\
        u(t) = u_\text{rest} + R w I_0 \left[ 1 - \exp \left(-\frac{t}{\tau_m}\right) \right] &\quad \text{if } 0<t<\Delta
    \end{cases}
\end{equation}
Where $w$ is the synaptic weight and $I_0$ is the current generated by a spike.
\\
The neuron spikes if $u \geq u_\text{threshold}$ ($u_\text{rest}<u_\text{threshold}$). If we choose $w$ such that $u(\Delta)=u_\text{threshold}$, the delay $b$, defined in section~\ref{section:network_architecture}, is given by $b=\Delta$, and $w$ is given by:
\begin{equation}
    \begin{aligned}
    &u(\Delta)=u_\text{threshold} \\
    \iff &u_\text{threshold} = u_\text{rest} + R w I_0 \left[ 1 - \exp \left(-\frac{\Delta}{\tau_m}\right) \right] \\ \iff &w = \frac{u_\text{threshold} - u_\text{rest}}{R  I_0 \left[ 1 - \exp \left(-\frac{\Delta}{\tau_m}\right) \right]}
     \end{aligned}
\end{equation}

\begin{figure}[htbp]
    \centering
    \includegraphics[width=0.3\textwidth]{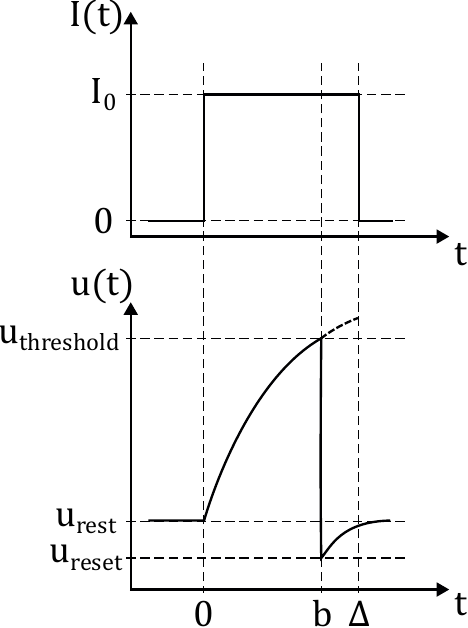}
    \caption{Neuron spike generation from an input current. Top: input current. Bottom: membrane potential of the neuron as a function of time.}
    \label{fig:spike_generation_LIF}
\end{figure}

If the neuron's membrane potential reaches the threshold, it is reset to a value $u_\text{reset}$. In this paper, we set $u_\text{reset} = u_\text{rest}$. The parameters and spike generation are shown in figure~\ref{fig:spike_generation_LIF}

\subsubsection{Recognition of a spiking motif of arbitrary length}
When a set of input neurons spike a retained motif, the network activates in a sequential manner. The input neurons spike the first sub-motif at time interval $I_f$. Input spikes propagate through the synapses with various delays, and will reach output neuron $o_f$ simultaneously at time $t_{if}=fI+a$. Neuron $o_f$ will therefore fire at time $t_{of}=fI+a + b$. Meanwhile, the input neurons generate spikes from the next sub-motif starting at time interval $I_{f+1}$. Input spikes and the spike from output neuron $o_f$ will reach synchronously the output neuron $o_{f+1}$, at time $t_{i(f+1)}=(f+1)I+ a$. Neuron $o_{f+1}$ will then fire at time $t_{o(f+1)}=(f+1)I+ a +b$. This sequence of activation of output neurons will go on until neuron $o_s$ will emit a spike, meaning that the sequence is recognized (cf. figure~\ref{fig:neural_architecture}a,c for visualization of sequential activation of spikes). If one output neuron does not spike, the next output neurons will not spike (except in some particular cases where parameters allow for noisy inputs, cf. eq.~\ref{eq:noisy_input_weight}), and the pattern will therefore not be recognized.

\subsubsection{Parallel recognition of multiple motifs of arbitrary length}
\begin{figure}[htbp]
    \centering
    \includegraphics[width=0.5\textwidth]{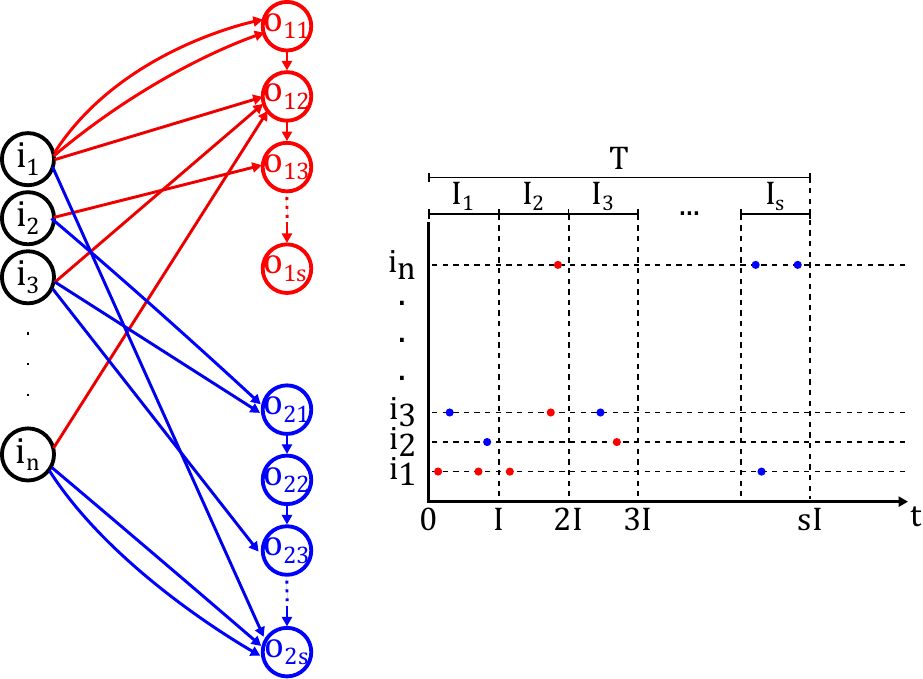}
    \caption{Neural network architecture for coding distinct motifs, here two motifs plotted respectively with blue and red spikes.}
    \label{fig:neural_network_two_learned_motifs}
\end{figure}

Since input neurons can generate several motifs, we may need to store more than one motif. For each motif we wish to memorize, a group of output neurons is connected to the input neurons. All sets of output neurons, encoding a motif, are independent from each other. A visualization of an architecture encoding two distinct patterns is given in figure \ref{fig:neural_network_two_learned_motifs}.

\subsection{Simulation of the network}
\label{section:Simulation of the network}
The network was simulated using Python with the PyNN Application Programming Interface (API)~\cite{Davison2008}, which provides a unified language. One advantage of using PyNN is that it allows for easy switching between different simulators, such as NEST, Brian2, or NEURON, without changing the code. This flexibility is beneficial for testing the network on different platforms and ensuring compatibility with various simulation environments but also as it allows to extend these simulations to neuromorphic hardware~\cite{Goltz2024}. For this work, we used PyNN with the NEST backend, which efficiently integrates the LIF neuron model described in equation~\ref{eq:LIF_membrane_potential} using optimized C++ code.

To test the recognition of a given set of known motifs, a network was built following the description in section~\ref{section:network_architecture}. For each motif, input neurons where connected to a group of output neurons, as represented in figure~\ref{fig:neural_network_two_learned_motifs}. The network was simulated using LIF neurons with fixed threshold and alpha-function-shaped post-synaptic current ($\text{IF\_curr\_alpha}$ on PyNN). To avoid long term effects of input spikes on the membrane potential of output neurons, the membrane time constant of the neurons was set to 1~ms such that neurons had fast dynamics. We also set a non zero refractory time for all neurons in order to avoid excessive spiking of neurons since the network was only composed of excitatory neurons. The initial, resting and reset membrane potential of the neurons where all set to -65mV, and the threshold was set to -50mV. 

The number of output neurons was defined with respect to the maximum synaptic delays and the duration of the encoded motif. For instance, 10 output neurons are required to memorize a 100~ms duration motif with maximum synaptic delays of 10ms. The synapses connecting input neurons to output neurons have constant weights and delays, as defined in section~\ref{section:network_architecture}. The synaptic weight of output to output neurons were 10 times higher than the weights of input to output neurons ($\gamma = 10$ in eq.~\ref{eq:weight_input_output_factor} and~\ref{eq:weight_output_output_factor})
Our code was partly compiled using Numba~\cite{lam2015numba}, and processed in parallel on multiple CPU's using Joblib. Reproducible code is available at \url{https://github.com/ThomasKronland/2026_Spiking_motifs_detection}.

\section{Results}
\label{section:results}

\subsection{Random synthetic spiking motifs} 
\label{section:Random_motifs_expe}
To evaluate the network's ability to recognize a stored motif in the presence of simultaneous motifs (i.e., noise), we generated random synthetic spiking motifs to serve as input. Each motif was constructed by specifying the minimum and maximum interspike intervals, the total motif duration, and the maximum number of spikes per neuron. For each neuron, spike times were drawn from a uniform distribution within the specified interval, with a minimum interval of 3~ms enforced to prevent excessive firing.

In all tests, the membrane time constant was set to 1~ms for all neurons and the refractory period was set to 1~ms for input neurons and 0.5~ms for output neurons, balancing the need to prevent excessive spiking while allowing motif detection in noisy conditions. All networks used a maximum synaptic delay of 10~ms. For each configuration, 40 independent trials were performed with randomly generated motifs to ensure statistical reliability.

\begin{figure}[htbp]
    \centering
    \includegraphics[width=.8\textwidth]{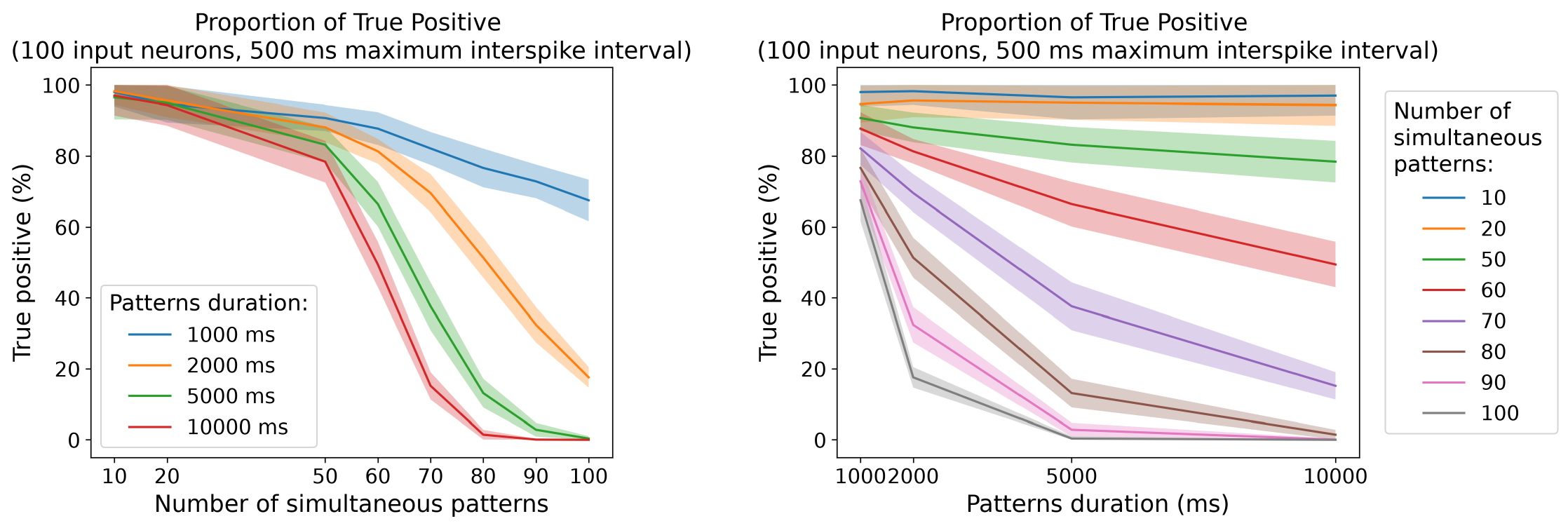}
    \caption{Results of the simulated network on simultaneous random motifs for various numbers of simultaneous motifs and motifs duration. The shaded bands give the standard deviation of the results. Each configuration was tested 40 times. Both figures correspond to the same simulation. We only display the proportion of true positive in this figure since the proportion of false positive is always null, and the proportion of false negative can be deduced from the proportion of true positive and eq.~\ref{eq:PFP_PFN_formula}.
    }
    \label{fig:random_motifs_simultaneous_results_num_duration}
\end{figure}

The first experiment evaluated the network's ability to recognize a single motif in the presence of multiple overlapping motifs, as a function of motif duration and the number of overlapping motifs. Here, the number of input neurons was fixed at 100, and the maximum interspike interval was set to 500~ms, resulting in a mean firing rate of approximately 4 Hz per neuron. Results are shown in Figure~\ref{fig:random_motifs_simultaneous_results_num_duration}. The proportion of correctly recognized motifs decreases as either the number of simultaneous motifs or the motif duration increases. This is expected, as more simultaneous motifs reduce the signal-to-noise ratio, and longer motifs duration increase the likelihood of spike conflicts at the output neurons. If an output neuron spikes at the wrong time due to noise, its refractory period may prevent it from firing when the actual motif occurs. Both effects contribute to an increase in false negatives, reducing the proportion of motifs successfully recognized.

\begin{figure}[htbp]
    \centering
    \includegraphics[width=1\textwidth]{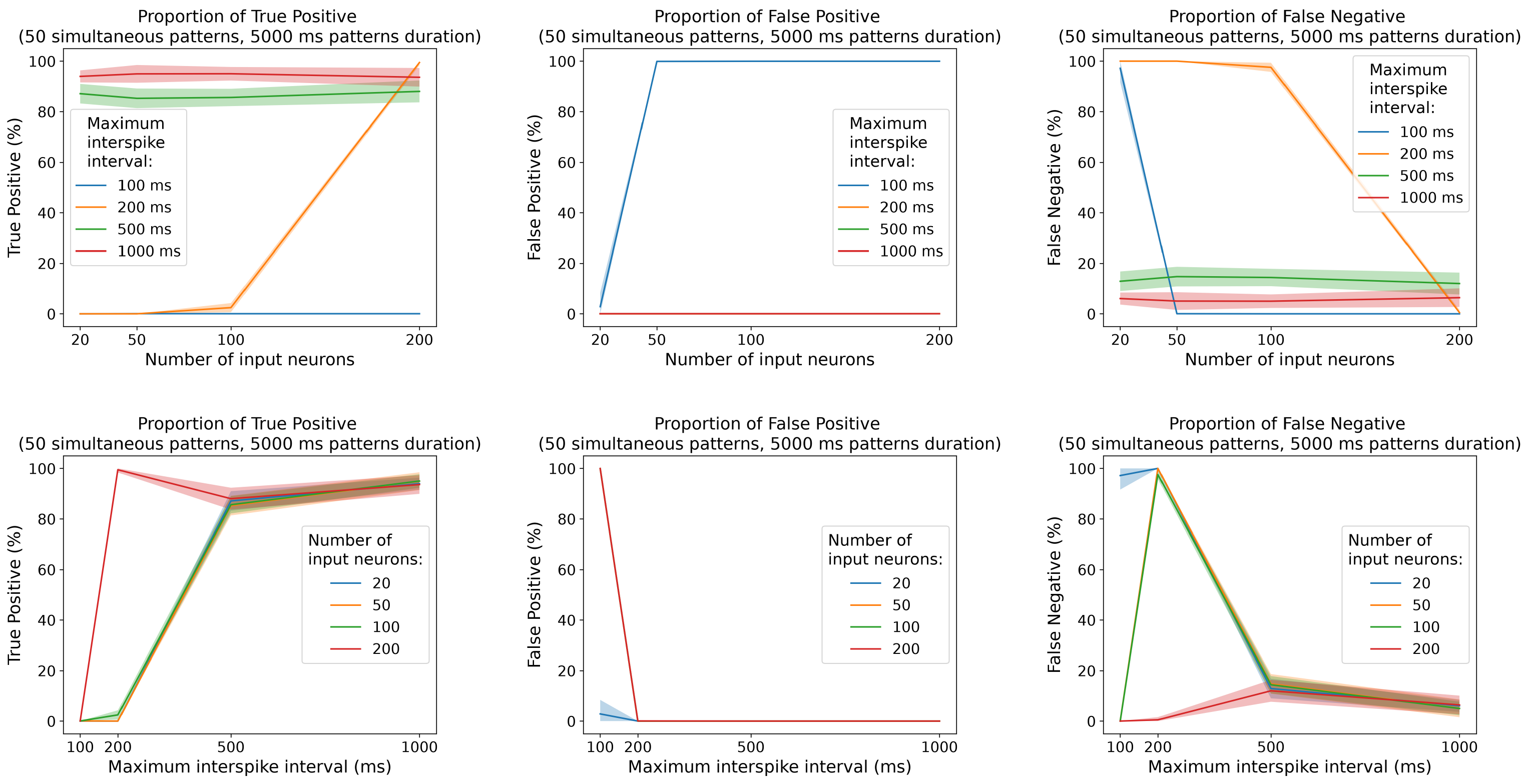}
    \caption{Results of the simulated network on simultaneous random motifs for various numbers of input neurons and maximum interspike interval. The shaded bands indicate the standard deviation across 40 trials per configuration. All figures correspond to the same simulation.}
    \label{fig:random_motifs_simultaneous_results_input_neurons_interspike_interval}
\end{figure}

The second experiment analyzed the network's ability to recognize a single motif in the presence of simultaneous motifs, as a function of the number of input neurons and the maximum interspike interval (which determines the firing rate). In this test, 50 motifs of 5000~ms duration were generated. Results are shown in Figure~\ref{fig:random_motifs_simultaneous_results_input_neurons_interspike_interval}.

The results indicate that the proportion of correctly recognized motifs increases with both the number of input neurons and the maximum interspike interval (i.e., as the firing rate decreases). Increasing the number of input neurons expands the space of possible motifs, reducing the likelihood that an output neuron is incorrectly activated by noise. False negatives typically occur when an output neuron spikes prematurely due to noise and is then unable to fire at the correct time due to its refractory period. Since output neurons are activated sequentially, a missed or mistimed spike in any output neuron prevents recognition of the entire motif.

Similarly, increasing the maximum interspike interval (lowering the firing rate) makes motifs sparser, which reduces the probability of spike conflicts and erroneous output neuron activation. At high firing rates, the proportion of false positives is elevated, but this decreases as firing rates drop. False negatives also decrease with sparser activation, as fewer spike conflicts occur. When input neurons are highly active, false positives dominate; as activity decreases, errors shift toward false negatives due to missed activations. Overall, correct recognition improves as input neuron activity becomes sparser.

In summary, these tests demonstrate that the motif recognition network is robust to noise, with optimal detection achieved for short motifs, a large number of input neurons, and sparse neuronal activation. An unexpected result in Figure~\ref{fig:random_motifs_simultaneous_results_input_neurons_interspike_interval} is that, for 200 input neurons, the highest recognition rate occurs at a maximum interspike interval of 200~ms. This phenomenon is not yet fully understood and suggests that optimal configurations may depend on the specific characteristics of the input motifs.

\subsection{Spiking Heidelberg Digits}
\begin{figure}[htp]
    \centering
    \includegraphics[width=.6\textwidth]{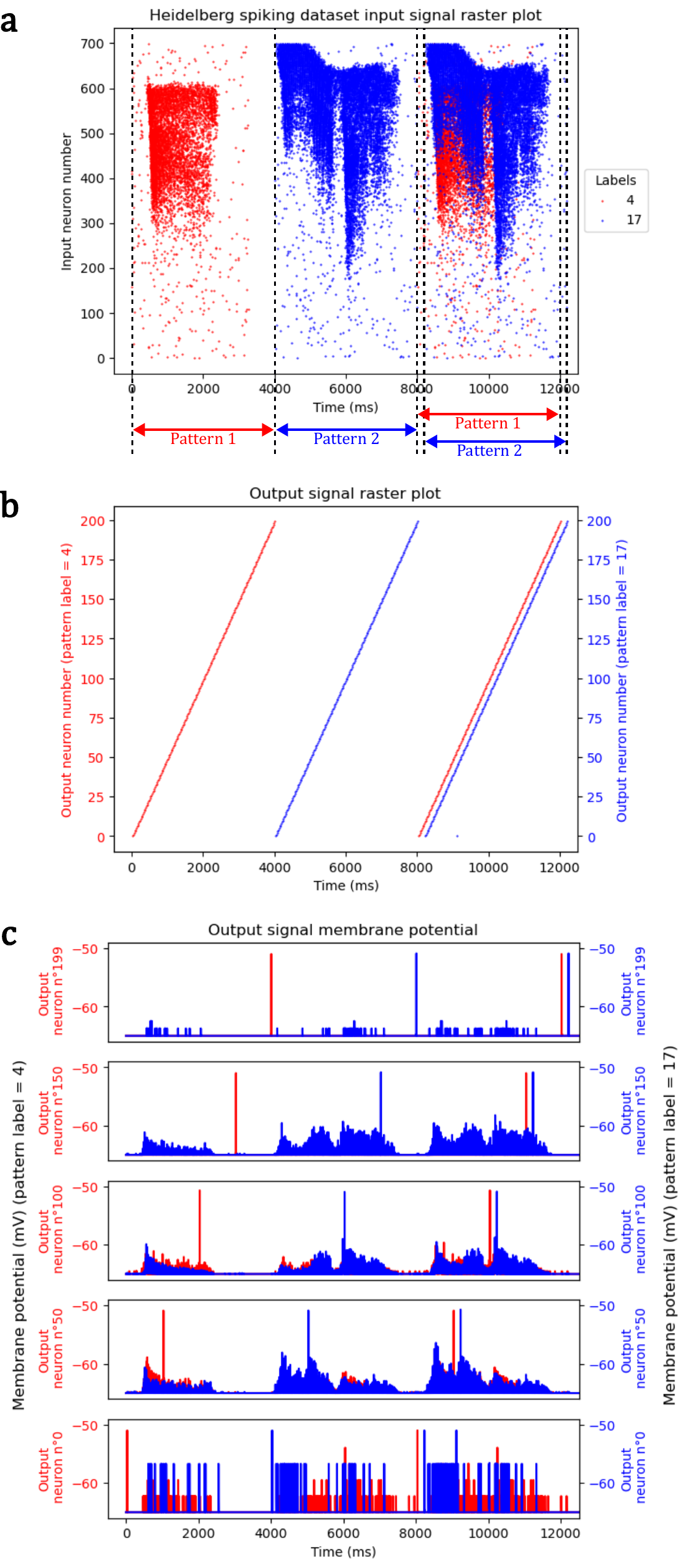}
    \caption{Simulation of the network on two motifs from the Spiking Heidelberg Digits (SHD) dataset. a) Two spiking patterns (in blue and red) corresponding to two different digits (4 and 17). To test the robustness of the NN, patterns were overlapped. b) Activation of the output neurons of each networks coding for each pattern. c) Membrane potentials versus time of 5 output neurons out of 200.}
    \label{fig:SHD_motif_recognition_results}
\end{figure}

The network was further evaluated using the Spiking Heidelberg Digits (SHD) dataset~\cite{cramer2020heidelberg}. This dataset consists of audio recordings of spoken digits (0–9) in both English and German, converted to spike trains via the Lauscher cochlear model. In figure~\ref{fig:SHD_motif_recognition_results}a, label ``4'' refers to the English digit 'four', and label ``17'' to the German digit 'sieben' ('seven'). The dataset was accessed using the Tonic Python library~\cite{lenz_gregor_2021_5079802}. To reduce simulation time, the spike data was downsampled by a factor of $10^{3}$. Since SHD data have microsecond resolution, this downsampling step leads to millisecond resolution of the data (using Tonic's \text{Downsample} transform), and the minimum interspike interval (ISI) was set to 5~ms to prevent excessive output neuron spiking.

For clarity, we standardized the motif duration to 4 seconds, even though the actual motifs varied in length. If a motif was shorter than 4 seconds (see motif 1 in figure~\ref{fig:SHD_motif_recognition_results}a), spikes simply propagated through the remaining output neurons without requiring additional input spikes. Each motif was encoded by connecting input neurons to a sequence of output neurons, with synaptic delays capped at 20~ms. Thus, a 4-second motif was represented by 200 output neurons (one per 20~ms interval). In figure~\ref{fig:SHD_motif_recognition_results}b-c, recognition of motif 1 was indicated by a spike from output neuron number 199, which corresponds to the final interval of the motif.

All neurons were configured with a refractory period of 2~ms and a membrane time constant of 1~ms. Connections followed the architecture described in Section~\ref{section:network_architecture}. For testing, the input signal consisted of: motif 1 (from 10~ms to 4010~ms), motif 2 (from 4010~ms to 8010~ms), and then both motifs presented simultaneously with an offset of 190~ms between the motif 1 (from 8020~ms to 12020~ms) and the motif 2 (from 8210~ms to 12210~ms). This offset allowed us to distinguish true positive motif recognition from false positives. It was chosen arbitrarily, but long enough to distinguish visually the recognition of both motifs by output neurons in Figure~\ref{fig:SHD_motif_recognition_results}b-c, while maximizing the overlap between both patterns, as seen in Figure~\ref{fig:SHD_motif_recognition_results}a.
Motif 1 (label 4) was detected at 4013.1~ms and 12023.1~ms, and motif 2 (label 17) at 8016.6~ms and 12216.6~ms, as indicated by spikes from output neuron number 199 in Figure~\ref{fig:SHD_motif_recognition_results}c. These detection times correspond to the end of each motif, confirming that the network accurately identified both the timing and identity of the motifs—even when they overlapped. 

\begin{figure}[htp]
    \centering
    \includegraphics[width=.4\textwidth]{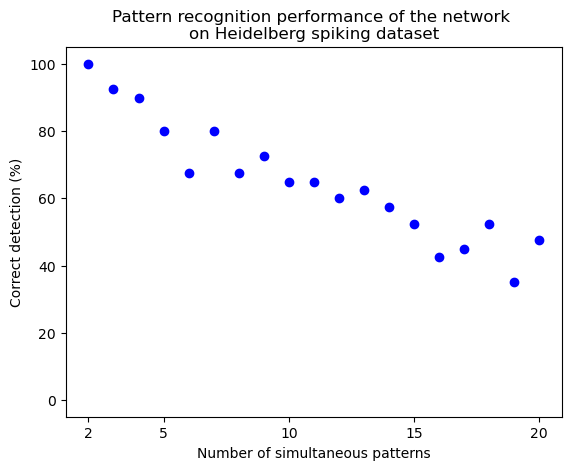}
    \caption{Performance of the network as a function of the number of simultaneous motifs from the Spiking Heidelberg Digits (SHD) dataset. For each condition, 40 tests were performed. Each point corresponds to the mean value over the 40 tests.}
    \label{fig:stats_test_SHD_motif_recognition}
\end{figure}

In addition, we performed similar tests with an increasing number of simultaneous motifs. In those tests, only overlapping motifs were tested, and motifs were offset from each other by 50~ms. The number of simultaneous motifs ranged from 2 to 20, and each condition was tested 40 times with random motifs from the SHD dataset. For each test, we considered that the recognition was successful only if all patterns were correctly detected. For instance, if one pattern was not recognized, the recognition was considered wrong. The mean proportion, over the 40 tests, of successful recognition for varying number of overlapping patterns is given in figure~\ref{fig:stats_test_SHD_motif_recognition}. This result demonstrates the network's robustness to noise and its ability to reliably recognize multiple motifs within complex input patterns. 

\section{Discussion}
\label{section:discussion}

This paper demonstrates that spiking neural networks with bounded synaptic delays can reliably recognize spiking motifs of arbitrary length. The proposed architecture is biologically plausible, reflecting key constraints observed in the brain. Notably, the network features a much higher number of synapses than neurons, mirroring biological circuits~\cite{Buzsaki2010}. Simulations show that motif recognition remains robust even when multiple motifs occur simultaneously, and that sparse neuronal activity further improves accuracy. This result is in line with experimental findings about cortical neurons typically firing less than one spike every 7 seconds~\cite{Lennie2003}. The robustness of our network to noise further supports the idea, introduced by Mainen and Sejnowski~\cite{mainen1995reliability}, that spike timings can remain reliable and precise even in the presence of input noise. Moreover, encoding information in precise spike timings, rather than firing rates, makes the network resilient to changes in stimulus intensity. For example, in the brain, sensory inputs are often encoded on a logarithmic scale, so variations in intensity affect firing rates but not the relative timing of spikes~\cite{hopfield1995pattern}. Interestingly, our architecture may be viewed as a continuous-time analogue of a McCulloch-Pitts network, where each neuron implements a simple logical operation and more complex computations emerge from their hierarchical combination~\cite{mcculloch1943logical}. In the coincidence detector regime, each output neuron behaves as such a threshold unit, detecting a local sub-motif, while the feedforward connections between successive output neurons ensure that the complete motif is detected only when all sub-motifs are detected in the correct order.

Despite these strengths, the network has several limitations. First, it currently includes only excitatory neurons. To prevent excessive spiking and false positives, we imposed a nonzero refractory period and a minimum interspike interval of 3~ms. Incorporating inhibitory neurons could help regulate activity through homeostatic mechanisms~\cite{Perrinet19hulk} and enable recognition of a broader range of motifs. Second, motif recognition depends on a single output neuron firing; if this neuron (or any in the output chain) fails, the motif is not detected—a challenge known as the 'grandmother cell problem'~\cite{gross2002genealogy}. Using populations of output neurons, rather than single cells, could improve robustness. Adding skip connections between output neurons may also allow partial motif recognition, even if one neuron fails to fire. Probabilistic neurons with stochastic escape rates could further enhance detection of noisy or incomplete motifs.

Future work should address several open questions. For example, the delays are set by construction and are not learned through repeated (unsupervised) presentation of the same pattern. However, the ability to memorize and recall a given pattern by a generic recurrent SNN has been recenty studied by \cite{perrinet2026working}. The network's memory capacity could be analyzed more formally, as initial results suggest high storage potential for simultaneous motifs. Recent advances show that Hopfield networks can achieve exponential memory capacity by optimizing synaptic weights~\cite{ramsauer2020hopfield}; combining weights and delays may further increase capacity in spiking networks. The impact of interspike interval distributions also merits investigation: while this study used uniform intervals, biological spike trains are often modeled as Poisson processes, which may affect recognition performance. Synaptic delays could be combined with other temporal mechanisms, such as adaptive firing or variable membrane time constants, to reduce the number of output neurons needed for motif encoding. Learning both delays and weights through supervised or unsupervised methods could enable motif classification; relevant synaptic learning rules have been proposed~\cite{grimaldi2023learning, perrinet2023accurate, hussain2014delay}. Additionally, while all spikes within a time interval were assigned equal synaptic weights in this study, some spikes may be more informative for motif recognition, for instance because they would be more selective to a motif, suggesting that heterogeneous weights could be beneficial. The network is well suited for processing asynchronous spiking motifs, such as those from neuromorphic sensors, and could be tested in real time on neuromorphic hardware. Finally, as discussed in~\cite{schlungbaum2023detecting} about the cocktail party problem, future studies could explore our network's ability to detect weak signals in the presence of periodic motifs.

\section{Materials and methods}
In section~\ref{section:Random_motifs_expe}, to systematically test robustness, motifs were designed to partially overlap rather than occur strictly simultaneously. This was achieved by introducing a fixed offset of 10~ms between the start times of each motif. For example, with three motifs of 100~ms duration, the first motif would span 10–110~ms, the second 20–120~ms, and the third 30–130~ms. Since the network typically recognizes a motif within 10~ms after its end, motif 1 was considered correctly recognized only if detection occurred between 110 and 120~ms. To ensure sufficient overlap, motif durations were chosen to be at least as long as the total offset (e.g., at least 30~ms for three motifs).

This approach allowed us to systematically assess the network's robustness to overlapping and noisy spike patterns.
Simulation results are reported as proportions of true positives, false positives, and false negatives in motif recognition. Specifically, the proportion of true positives $P_{TP}$ is defined as:
\begin{equation}
    P_{TP} = \frac{TP}{TP + FP + FN},
\end{equation}
where $TP$ is the number of correctly recognized motifs (true positives), $FP$ is the number of incorrectly recognized motifs (false positives), and $FN$ is the number of missed motifs (false negatives). We do not consider the number of correctly ignored motifs (true negatives) here, since this number would be too large.
Similarly, the proportions of false positives $P_{FP}$ and false negatives $P_{FN}$ are given by:
\begin{equation}
\begin{aligned}
    P_{FP} &= \frac{FP}{TP + FP + FN}, \\
    P_{FN} &= \frac{FN}{TP + FP + FN} = 1 - P_{FP} -  P_{TP}
\label{eq:PFP_PFN_formula}
\end{aligned}
\end{equation}

\section*{Acknowledgments}
TKM, SV and LP received funding from the ANR project number ANR-20-CE23-0021 (``\href{https://laurentperrinet.github.io/grant/anr-anr/}{AgileNeuRobot}'') and LP from the french government under the France 2030 investment plan, as part of the Initiative d’Excellence d’Aix-Marseille Université – A*MIDEX grant number AMX-21-RID-025 ``\href{https://laurentperrinet.github.io/grant/polychronies/}{Polychronies}''. TKM acknowledges the Institut de Recherche sur les Composants logiciels et matériels pour l’Information et la Communication Avancée (IRCICA, CNRS, Université de Lille, France) for their kind welcome during the completion of this work. For the purpose of open access, authors have applied a CC BY public copyright licence to any Author Accepted Manuscript version arising from this submission.

\bibliographystyle{unsrt}
\bibliography{spikes}

\end{document}